\documentclass[reprint,amsmath,amssymb,aps]{revtex4-1}

 \usepackage{graphicx}

\begin{document}

\title{Simulations of Particle Impact at Lunar Magnetic Anomalies and Comparison with Spectral Observations}

\author{Erika Harnett}
\email{eharnett@uw.edu}
\affiliation{Department of Earth and Space Science, University of Washington,Seattle, WA 98195-1310, USA}
\author{Georgiana Kramer}
\affiliation{Lunar and Planetary Institute, 3600 Bay Area Blvd, Houston, TX 77058, USA}
\author{Christian Udovicic}
\affiliation{Department of Physics, University of Toronto, 60 St George St,Toronto, ON M5S 1A7, Canada}
\author{Ruth Bamford}
\affiliation{RAL Space, STFC, Rutherford Appleton Laboratory,Chilton, Didcot Ox11 0Qx, UK}

\date{\today}

\begin{abstract}
Ever since the Apollo era, a question has remained as to the 
origin of the lunar swirls (high albedo regions 
coincident with the regions of surface magnetization). 
Different processes have been proposed for their origin. 
In this work we test the idea that the lunar swirls have a 
higher albedo relative to surrounding regions 
because they deflect incoming solar wind particles that would 
otherwise darken the surface. 
3D particle tracking is used to estimate the influence of five 
lunar magnetic anomalies on incoming solar wind. 
The regions investigated include Mare Ingenii, Gerasimovich, 
Renier Gamma, Northwest of Apollo and Marginis. 
Both protons and electrons are tracked as they interact with 
the anomalous magnetic field and impact maps are calculated. 
The impact maps are then compared to optical observations and 
comparisons are made between the maxima and minima in 
surface fluxes and the albedo and optical maturity of the regions. 
Results show deflection of slow to typical solar wind particles on a 
larger scale than the fine scale optical, swirl, features. It is 
found that efficiency of a particular anomaly 
for deflection of incoming particles does not only
scale directly with surface magnetic field strength, but 
also is a function of the coherence 
of the magnetic field. All anomalous regions can also produce 
moderate deflection of fast solar wind particles. 
The anomalies' influence on $\sim$ 1 GeV SEP
particles is only apparent as a slight modification of the incident velocities. 
\end{abstract}



\maketitle

\section{Introduction}  

Lunar swirls are high albedo regions on the lunar surface which appear to correspond to surface magnetic anomalies. 
(See reviews by \cite{Lin1988} and \cite{Blewett2011}).  
While the origin of the lunar swirls is not yet resolved, one of the main theories is that 
the anomalous magnetic field deflects incoming solar wind, which would otherwise impact the surface and chemically weather, 
or darken, the lunar regolith through the creation of nanophase iron (\cite{Hood1980}, \cite{Hood1989},\cite{Hood2001},
\cite{Hood2008}, \cite{Kramer2011a}, and \cite{Kramer2011b}). 
These incoming particles may be completely 
deflected away from the surface or they may be deflected to other regions on the surface. It is thought that the 
dark lanes, regions of very low albedo adjacent to swirls, may correspond to locations of enhanced particle flux 
and, thus weathering, due to nearby particle deflection (For a more in-depth discussion see \cite{Kramer2011a}). 

The idea that the lunar magnetic anomalies may deflect incoming solar wind was first proposed during the Apollo era
to explain compression of the anomalous magnetic field, or amplifications of the magnetic field
 near the limb (called ``limb compression'') as observed from orbit [\cite{Dyal1972} and \cite{RussellLicht1975}]. 
Observations by Lunar Prospector gave the first conclusive evidence 
that the magnetic anomalies could deflect the solar wind, forming mini-magnetospheres [\cite{Lin1998}]. Subsequent 
observations by Lunar Prospector [e.g. \cite{Halekas2006} and \cite{Halekas2008}], Nozomi [\cite{Futaana2003}],
SELENE/Kaguya [e.g. \cite{Saito2008a}, \cite{Wieser2010} and \cite{Lue2011}], Chandrayaan-1 [\cite{Holmstrom2010}],
and Chang'E-2 [\cite{Wang2012}] further confirmed that the magnetic anomalies can not only 
deflect incoming solar wind particles but also modify the distribution [\cite{Saito2012}].

Different processes may be involved in deflecting incoming particles, and the relative importance of each process
will be a function of the surface magnetic field strength and the scale size of the anomalous region. 
One process for deflecting particles is called magnetic mirroring. 
Charged particles move with both a spiral motion perpendicular, and 
parallel to the magnetic field. If the particles move into a region with magnetic field that increases in 
magnitude (or converges), the kinetic 
energy of the particle parallel to the magnetic field will be converted into kinetic energy perpendicular to the magnetic field.
Eventually, if the magnetic field magnitude is strong enough for a given incident kinetic energy, all of the energy
will be converted to perpendicular to the magnetic field, and the particle will be reflected. 

The initial observations of deflection around the Imbrium antipode region by Lunar Prospector [\cite{Lin1998}] 
suggested that if the anomalous region 
is large enough, the incident solar wind plasma has a collective behavior, and the plasma behaves 
in a fluid manner (see also \cite{Halekas2006}). The dynamic pressure of the incident plasma was 
balanced by the magnetic pressure of the anomaly, slowing the plasma, producing signatures of a shock region forming, 
and mini-magnetosphere, around the anomaly. 
Subsequent observations of the solar wind interacting with a wider range of anomalous regions for a variety of solar
wind conditions has lead to a complex picture of the interaction. Observations by Chandrayaan-1, in the vicinity of both
strong and weak anomalies [\cite{Lue2011}], revealed a plasma interaction in which the electrons behaved in a fluid manner, 
while the protons became demagnetized. This would lead to charge separation and the generation of an ambipolar electric field,
which would act to both accelerate electrons and slow protons. More recent observations have also confirmed the presence
of density cavities around magnetic anomalies (e.g. \cite{Saito2012}, \cite{Wieser2010}). While these observations help
refine our understanding of the physics governing the formation of mini-magnetospheres, there is still uncertainty with 
regard to how the plasma deflection seen above the surface may be connected to the high albedo swirls seen on the 
surface.

Particle tracking by \cite{Hood1989} suggested that the Reiner Gamma region, modeled as a collection of dipoles,
has sufficient magnetic field strengths to deflect solar wind ions.  Particle tracking was also employed by 
\cite{Harada2014} in order to investigate why backscatter ENAs were typically observed by Chandrayaan-1 over the 
magnetic anomalies in the southern hemisphere while in the solar wind but not in the terrestrial plasma sheet.
They found that when a nearly monoenergetic, monodirectional population of protons (analogous to the solar wind) 
interacted with a subsurfacemagnetic field, a density cavity formed near the surface. Conversely, when an isotropic,
Maxwellian distribution of particles (analogous to plasma sheet protons) interacted with the subsurface dipole, the 
density cavity near the surface was greatly reduced in size, due to the increased spread in incident particle 
directions. 

Self consistent 2.5D fluid and particle simulations of the solar wind interacting with a small 
dipole on the surface of the Moon by \cite{Harnett2002} showed that neither the
small scale size of a magnetic anomaly nor kinetic effects from the different behavior of ions and electrons
prevent a shock-like region from forming. Additional fluid simulations 
modeling the anomaly as a collection of dipoles [\cite{Harnett2003}] showed that the nature of the 
shock would be highly dependent on the structure of the magnetic field and the orientation of the IMF. 3D Hall-MHD
simulations by \cite{Xie2015} have produced a similar result.
\cite{Kallio2012} used the combination of a 3D hybrid model and a 1D Particle-in-a-Cell (PIC) model to look at the
kinetic effects of the solar wind interacting with a dipole field both locally (PIC) and globally (hybrid). They 
found that the central portion of the dipole field could completely block access of protons to the surface while
the surrounding regions showed enhanced density and flux. \cite{Poppe2012b} used a 1.5D PIC model to investigate
the solar wind interacting with cusp-like structures that may be present at some magnetic anomalies, and possibly
be co-located with the dark lanes. They looked at a variety of surface magnetic field strengths and found significant
modification of the interacting plasma relative to the incoming distribution. They saw ion deceleration and electron
acceleration similar to that observed by Kaguya [\cite{Saito2012}]. \cite{Bamford2012} used a vacuum chamber to 
look at plasma incident on two different dipole magnetic fields (a strong and a weak magnet) 
and compared the deflection with that seen by 
theory and satellite observations. They found that a shock-like structure formed around both magnetic fields, and
that the general shape of the structure that formed around both magnets was similar, even though the structure
around the weak magnet was considerably smaller. 

In this work, we present results from 3D particle tracking studies following the interaction of protons and electrons 
using 3D vector magnetic fields models of five different lunar magnetic anomalies, generated from satellite observations 
[\cite{Purucker2010}]. This work is unique in that it looks at how solar wind particles may interact with realistic 
anomalous magnetic fields over an extended region and attempts to correlated the particle response with observations
of the lunar swirls. The results presented here, begin to address the open question of why not all magnetic 
anomalies are associated with lunar swirls and why some anomalies with weak or moderate magnetic field strengths have
more extensive swirl regions than other, comparatively stronger magnetic anomalies. 
The five magnetic anomaly regions investigated are Mare Ingenii, Reiner Gamma, Gerasimovich, Marginis, and 
Northwest (NW) of Apollo,.
The first three were selected for study as they are classified among the strongest anomalies and 
have observable swirl characteristics [\cite{Blewett2011}]. Marginis was selected as it is classified 
as a weak anomaly but is co-located with a complex
swirl pattern, similar in general nature to the swirls at Mare Ingenii, a strong anomaly region. NW of Apollo
was selected as it is one of the moderate anomalies but does not have an easily identifiable swirl
region. The goal for NW of Apollo is to test the ability of the particle tracking to help guide the search 
for swirl regions. 

Particle tracking was employed as a way to both investigate particle deflections at a wide 
selection of anomalous regions for a variety of incident particle energies in a feasible time frame, 
and assess how effective just the anomalous magnetic field is alone in deflecting incoming particles. 
As discussed above, it is difficult to resolve from observations
alone the relative importance of anomalous magnetic field, ambipolar electric fields and kinetic effects in the 
formation and structure of mini-magnetospheres. This study allows for the quantification of the effect of the 
anomalous field alone in influencing the incident plasma.  \cite{Bamford2015}, presents results of fully 
self-consistent particle simulations for the Reiner Gamma region, assuming a dipole magnetic field. A comparison
of the results in this paper with those in \cite{Bamford2015} quantify how much the incident 
plasma in additionally influenced by the development of charge separation and the resulting 
ambipolar electric field. That paper also discusses 
the similarities and differences with the results from 3D PIC simulations by \cite{Deca2015}. 

As part of this work, impact maps for each simulated anomalous 
region are generated and co-located with both optical and maturity observations of the same regions. 
The results presented here focus
on solar wind regime incident particle energies, but do look at the possibility for 
deflection of solar energetic
particle (SEP) events associated with solar activity. Protons were selected as 
the ion species as solar wind hydrogen
is considered responsible for the creation of nanophase iron, which causes 
darkening and reddening of the surface
spectra as a soil matures (\cite{Kramer2011a},\cite{Kramer2011b}).

\section{Method}
\subsection{Particle Tracking}

Simulated anomalous magnetic field were generated at www.planet-mag.net/index.html, using the Correlative model 
described in \cite{Purucker2010}. The model magnetic fields were generated from observations, typically with 
passes separated by \(1^{o}\), at altitudes down to 30 km.  The model vector magnetic fields were 
generated at a given altitude, over a range of latitudes and longitudes appropriate for each case. 
The resolution of the model magnetic field in latitude and longitude ranged from \(0.1^{o}\) to \(0.15^{o}\). 
This was selected as it would create a magnetic field model with a resolution similar to the optical
images ($\sim$ 100m/pixel). In reality, the resolution of the magnetic field model scales with the lowest
altitudes of the observations made by Lunar Prospector (> 10s of km).

Planes of magnetic field were generated at 0.5 km slices up to approximately 100 km. The upper bound for each case was 
determined by where the anomalous magnetic field could not be distinguished from the background of $\sim$ 0.1 nT.
The constant altitude slices were stitched together to make a three dimensional grid with vector magnetic fields
at each grid point.  The simulated region size varied for each case, 
but for each included the central anomalous region plus several degrees surrounding. This allowed incident particles to be 
deflected without undergoing an interaction with the simulation side boundaries. Some of the magnetic anomalies 
studied form extended regions. In these cases, only the central, peak magnetic field region was of interest.
The basic characteristics for the five regions studied are given in Table ~\ref{anomalies}.
The total magnetic field simulated included four cases: just 
the anomalous magnetic field, and the anomalous magnetic field plus a superposition of three different 
interplanetary magnetic fields: \(\rm{B}_{\rm{vertical}} = \pm 2 \rm{nT,}\rm{B}_{\rm{horizontal}} = 2 \rm{nT}\), 
where horizontal and vertical are relative to the surface with the magnetic anomalies.

\begin{table*}[t]
\centering
\begin{tabular}{|l||l|l|l|lc|}                      \hline
Anomaly         &   Latitude      &   Longitude     &  Peak Field Strength & Peak Field Strength &\\ 
                &                 &                 &  30 km (nT)          & surface (nT)        &\\ \hline
Mare Ingenii    &  33.5\(^{o}S\)  & 160\(^{o}E\)     &  20                  &   75                &\\ \hline
Reiner Gamma    &  7.5\(^{o}N\)   & 302.5\(^{o}E\)   &  22                  &   56                &\\ \hline
Gerasimovich    &  21\(^{o}S\)    & 236.5\(^{o}E\)   &  28                  &   72                &\\ \hline
NW of Apollo    &  25\(^{o}S\)    & 197.5\(^{o}E\)   &  12                  &   49                &\\ \hline
Marginis        &  16\(^{o}N\)    & 88\(^{o}E\)      &  6                   &   29                &\\ \hline
\end{tabular}
\caption{Values in columns 2-4 taken from \cite{Blewett2011}. Magnetic field strengths in column 4 are 
estimates at 30 km altitude from satellite observations. Surface magnetic field strengths in column 5
are from the \cite{Purucker2010} model used as input for the particle tracking.}
\label{anomalies}
\end{table*}

For the particle tracking studies, 400,000 non-interacting protons or electrons were launched at the magnetized surface 
for the variety of total magnetic field configurations. Initial locations, within the launch region, were randomly 
assigned. Particle trajectories were computed using the Lorenz force law
until all the particles either impacted the surface or left the simulation area. The particles were not forced to 
move only on grid points where the magnetic field was defined, but rather could have arbitrary locations within the 
simulation region. The magnetic field at each particle location was computed in a weighted fashion from the 
magnetic fields defined at nearest grid points. 
For all the anomalous regions simulated, the particle flux maps showed little difference between the cases with 
an interplanetary magnetic field (IMF) and the case without an IMF, and little difference among the different 
IMF cases. The results shown here will focus on the cases with the IMF equal to \(\rm{B}_{\rm{vertical}} = -2 \rm{nT,}\). 
For all cases, runs were also completed that kept all aspects of the parameter space the same, except the anomalous
magnetic field was removed. This was done to allow for an estimate of the level of uncertainty in the results when
calculating variations caused by the anomalous magnetic field, for a specific case, and verify that any features
seen in the density and flux maps are in fact associated with modification by the anomalous magnetic field and 
not an artifact of non-randomness in the initial random location of the particles. 

The velocity distributions for the baseline cases have a 
mean of 200 km \(\rm{s}^{-1}\), and a Gaussian distribution with a thermal speed of 75 km \(\rm{s}^{-1}\). 
This speed represents either a slow solar wind speed or a high speed flow in the terrestrial magnetotail. 
Plasma from two sources will impact the lunar surface and contribute to weathering - the solar wind composed 
primarily of hydrogen ions, and terrestrial plasma sheet plasma composed of varying concentrations of hydrogen and 
oxygen ions. The plasma in the terrestrial plasma sheet will typically have a much broader range of thermal speeds
than the solar wind. The narrow thermal distribution was retained for this slowest speed to facilitate 
comparison among cases. 

For each case, accumulated densities and accumulated fluxes
(i.e. density of the impacting particles at a grid point times the average speed of the particles
at that grid point) for the five anomaly regions were calculated to compare with the optical and (if available) OMAT
images. Swirl mappings from the observations are compared to the density and flux maps. 
It is important to note that as these are results from particle tracking, the proton impact maps 
represent a maximum impact model. In all cases the trajectories for electrons were also determined (but not all are
shown). As the electrons are much more easily deflected by the anomalous magnetic field, many of the incident electrons do 
not impact the surface, and instead are completely reflected back, away from the surface. 
The electron simulations were run until all of the particles initially launched towards the surface either impacted 
the surface or were deflected away from the surface (either towards one of the simulations walls or back upstream).

Although both hydrogen ions and electrons were tracked in our simulations, only the trajectories of hydrogen ions 
were used for comparison with optical imagery as only they can impact with sufficient energy to both break bonds 
and be utilized as the reducing agent to create nanophase iron.
\cite{Reed1971} indicates that the energy required to break the FeO bond is $\sim$ 3-5 eV over a range of several 
Kelvin to a couple thousand Kelvin. 
This energy is equivalent to the kinetic energy of a 30 km \(\rm{s}^{-1}\) 
proton. \cite{Velbel1999}, on the other hand, indicates that an energy of 50 eV is required to break the FeO bond
at approximately 300 K. This energy is equivalent to a 100 km \(\rm{s}^{-1}\) proton.
Realistically though, some percentage of incident protons will scatter off other 
minerals within the regolith, loosing energy, before they encounter an FeO molecule, and not all of the kinetic
energy from the incident proton will necessarily be transferred to breaking the bond. For comparison, 
250-300 km \(\rm{s}^{-1}\) is the bulk speed at which it is estimated that protons, with a temperature of 5-10 eV, 
will produce a maximum sputtering yield from the lunar surface (\cite{Poppe2014} and references therein). 
The case of 200 km \(\rm{s}^{-1}\) (or a proton with a kinetic energy of $\sim$ 200 eV) is therefore treated in this
paper to be near the real minimum energy needed to weather the lunar regolith but not necessarily produce sputtering. 
Knowing an exact minimum in a realistic setting would require more extensive modeling and experiments to determine.

Additional cases for the \(\rm{B}_{\rm{vertical}}\) = -2 nT case were run with a 
mean proton velocity of 400 km \(\rm{s}^{-1}\) (typical solar wind velocity), 
2000 km \(\rm{s}^{-1}\) (i.e. fast solar wind) or 40,000 km \(\rm{s}^{-1}\) (i.e. $\sim$ 1 GeV 
SEPs - relativistic effects not included). This IMF case was also run for incident 
electrons with a mean velocity of 200 km \(\rm{s}^{-1}\) or 400 km \(\rm{s}^{-1}\) at each anomaly.
Total densities and fluxes at the surface were computed by distributing the particles, in a weighted manner, 
on to a grid with the same resolution as the magnetic field data, and summing over the collected particles. 
Densities and fluxes were normalized so that the super-particle density in the initial launch region 
corresponded to 5 particles \(\rm{cm}^{-3}\), nominal solar wind densities at 1 AU.

\subsection{Swirl Identification and Mapping}

Mapping and spectroscopic analysis of the swirls used data from Clementine, Lunar Reconnaissance Orbiter (LRO) 
cameras and Global Lunar Digital terrain model (GLD100) [\cite{Scholten2012}].
This data was supplemented, during analysis, with OH abundances measured by the  Moon Mineralogy Mapper (M$^3$)
on Chandrayaan-1 to ensure consistency with previous results [\cite{Kramer2011b}]. 
Clementine ultraviolet-visible (UV-VIS) and near-infrared (NIR) 
DIMs [\cite{Nozette1994}] were resampled to 100 m/pixel, combined into seamless 11-band images cubes, and the 
empirically-derived correction factors of \cite{Lucey2008} 
(USGS Clementine NIR global mosaic, available at http:// astrogeology.usgs.gov/Projects/ClementineNIR/) were applied. 
The cubes were then  were processed, mosaiced, projected (simple cylindrical), and co-registered  
to match the magnetic field maps using the Environment for Visualizing Images (ENVI). 
With the exception of Marginis, the basemap, upon which 
the swirls are outlined for each swirl region (Figures \ref{ingenii}g, \ref{reiner}g, 
\ref{geras}g, \ref{nw}e), is a simulated true (sim-true) color image generated from 
Clementine data (red = 900 nm, green = 750 nm, blue = 415 nm). The Clementine data for Marginis suffered 
too many gaps in coverage. Therefore, the basemap for Marginis (Figure \ref{marg}e) used data from 
LRO's Wide Angle Camera (WAC).

Lunar swirls can be difficult to unambiguously identify, and more so to assign boundaries to, owing to their often 
diffuse nature, range in shapes and sizes, and contrast against the terrain on which they occur. Although the swirls 
are high albedo, so usually easily distinguished against the dark background of the maria, they tend to blend into the 
surrounding terrain when they overly bright highlands material. Therefore, in addition to band albedo images, we 
generated spectral parameter (SP) images from both Clementine and M$^3$ data to facilitate mapping the swirls and to identify 
broad spectral characteristics of the surface that coincide with the modeling results. 
A spectral parameter utilizes the spectral features that are specific to an attribute of scientific 
interest in an algorithm in order to accentuate that attribute. For example, the wavelength and depth of absorption 
features can be used to identify specific minerals; the peak albedo and slope of the entire spectral continuum can be 
used to estimate maturity. A spectral parameter image is created by applying such an algorithm to each pixel of a 
spectral image to derive the spatial context of the desired feature, or parameter. Several different SP maps were 
needed to outline the swirls, because different SPs accentuate different spectral attributes of swirls, as well as other 
geologic features that share that spectral attribute. No single SP has been found that is unique to the swirls. For example, 
the swirls are optically immature so appear bright in an optical maturity parameter (OMAT) map [\cite{Lucey2000b}], 
however, so does ejecta from fresh impact craters. In addition, \cite{Kramer2011b} showed that the swirls are 
depleted in OH relative to their surroundings using OH abundance maps generated from M$^3$ mosaics. This makes the 
relative OH abundance parameter a strong swirl identifier, although not absolute, since OH abundance also varies as a 
function of the angle between the Sun and the surface (time of day, slope, latitude, etc.) 
[\cite{Pieters2009, McCord2011}]. However, considered collectively the various SP maps aid in 
identifying swirls better than any individual image.

\section{Particle Tracking Results}

\subsection{Mare Ingenii}  

\begin{figure*}[t]
\begin{center}
\includegraphics[height=0.8\textheight]{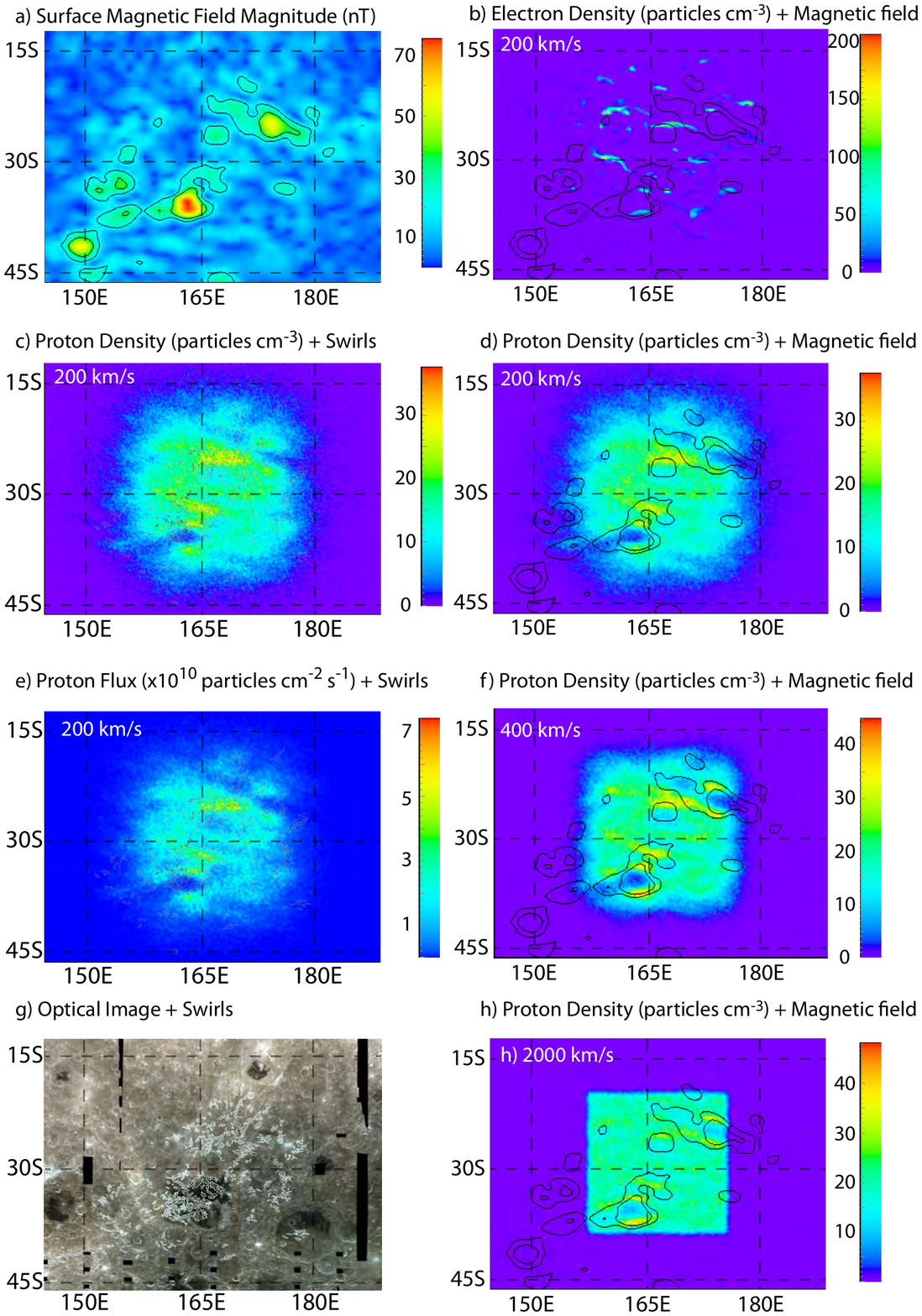}
\caption[Ingenii Maps]
{\textbf{Mare Ingenii} - The (a) surface magnetic field at Mare Ingenii. Also shown are the (b) surface electron density
with magnetic field contours overlayed, 
(c) surface proton density with swirls (in black) overlayed, (d) surface proton density with magnetic field overlayed, 
and (e) surface proton flux with swirls (in black) for the case of particles with a 
mean incident velocity of 200 km \(\rm{s}^{-1}\) and an IMF of \(\rm{B}_{\rm{vertical}} = -2 \rm{nT}\). 
The swirls are mapped from the (g) optical image and shown in cyan. The surface proton densities and magnetic field contours
are also shown for the cases of incident protons with velocities equal to 
(f) 400 km \(\rm{s}^{-1}\) and (g) 2000 km \(\rm{s}^{-1}\). 
The magnetic field contour lines show anomalous magnetic field magnitudes of 20 nT and 35 nT.
Animations of figure components can be found at: http://earthweb.ess.washington.edu/eharnett/papers/LunarImpactMovies/}
\label{ingenii}
\end{center}
\end{figure*}

\begin{figure}[t]
\begin{center}
\includegraphics[width=0.45\textwidth]{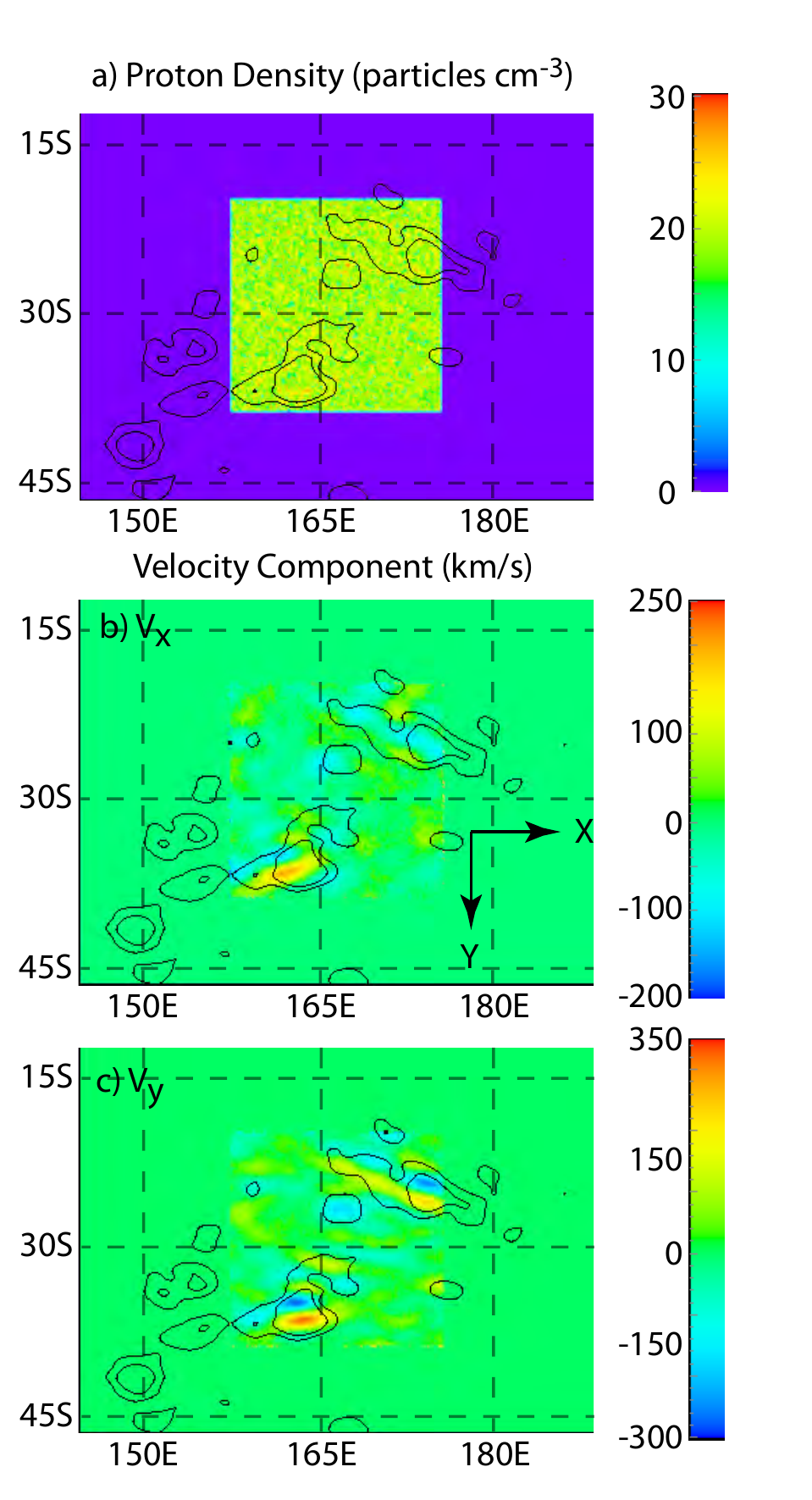}
\caption[Ingenii Velocity]
{\textbf{Mare Ingenii} - Surface densities for protons with a mean incident velocity of (b) 40,000 km \(\rm{s}^{-1}\).
Also shown are the average components of the velocity of particles when they impact the surface (b) along the
East-West direction and (c) along the North-South direction at Mare Ingenii. Arrows indicate the positive
directions for each component. The magnetic field contour lines in all three figures 
are identical to those in ~\ref{ingenii}.}
\label{ingenii_seps}
\end{center}
\end{figure}

Figure ~\ref{ingenii} shows  the results for the swirl region at Mare Ingenii. 
The swirls are outlined in light cyan based on optical imagery in Figure ~\ref{ingenii}g. These outlines are shown, 
in black, overlain on a proton density map (Figure ~\ref{ingenii}c), and an accumulated proton flux map 
(Figure ~\ref{ingenii}e), for the baseline case to compare model results with the locations of the high concentration of swirls.
The strongest surface magnetic field 
(\(\sim 35^{o}\)S and \(160^{o}\)E, Figure ~\ref{ingenii}a) is seen to correspond with low particle densities 
at the surface (Figure ~\ref{ingenii}d) and a void in the flux (Figure ~\ref{ingenii}e). Surrounding the 
void regions, for both density and flux, are regions with enhanced density and flux. The flux and 
density at the surface when no magnetic field is present is approximately \(2.5 \times 10^{10}\) particles 
\(\rm{cm}^{-2}\) \(\rm{s}^{-1}\) and 18 particles \(\rm{cm}^{-3}\). 
The reduction in the flux in the void regions is primarily due to the reduction in density. And while the speed 
of the particles that do impact in and around this portion of the anomaly is reduced by approximately 5\%, 
the velocity provides a more complete picture. In the central portion of the anomaly,  
the component of the velocity perpendicular to the surface decreases by 
approximately 75 - 100 km \(\rm{s}^{-1}\) while the magnitude of the components parallel to the surface increase 
from approximately zero to 75 km \(\rm{s}^{-1}\) (as compared to when no magnetic
field is present). 

Central to the void region at \(\sim 35^{o}\)S and \(160^{o}\)E
is the portion of the lunar swirl with the highest optical albedo (Figure ~\ref{ingenii}g). 
This corresponds with the brightest, bluest (flat spectral continuum), and most optically immature swirl surface at Ingenii.
Both to the north and south of this void are regions of enhanced surface flux and density. It is harder to
correlate these regions of enhanced flux with dark lanes purely from the optical image alone in part because 
these locations are coincident with the rims of the mare-filled craters Thompson and Thompson M,
which, being rich in the plagioclase, cannot darken like the dark lanes on the maria.
The simulations begin to describe the interactions that occur between the particles and the magnetic field
that are pattern manifested as complex patterns of bright and dark on the surface. This is demonstrated by
comparing the simulation results and swirl outlines within Thompson Crater, where the void in the proton flux
is coincident with a group of swirls. 
Unfortunately, the simulations stops short of describing the intricacy of the dark lanes observable in the optical 
images due to variations from high to low flux/density regions with scale sizes much larger than the scale sizes of
the swirls. This is a consequence of the coarser resolution of the model magnetic field data, as the resolution of 
the optical image in Figure ~\ref{ingenii}g (100m/pixel) is much smaller than the resolution of the observed magnetic field
(> 10s of km). 

The highest surface impact density and flux is associated with the region of moderate magnetic field around 
\(\sim 25^{o}\)S and \(170^{o}\)E. This peak magnetic field in this region is about 70\% that of the anomaly 
at \(\sim 35^{o}\)S and \(165^{o}\)E, but it is less localized. It is a region in which the components of the
magnetic field both perpendicular and parallel to the surface are fairly uniform over the whole extended region.
This means that particles incident from above this whole region will be deflected around it, translating into a
higher density, as particles accumulate from over a more extended than that for the stronger anomaly. The region 
of highest density and flux is co-located with a darkened region, surrounded by swirls (Figure ~\ref{ingenii}g).

Figures ~\ref{ingenii}b and ~\ref{ingenii}d show a comparison of the density impact maps for electrons and 
protons, with the same incident velocity. 
On the order of 80\% of the electrons were deflected away from the surface, while, at most, a few percent of the protons
did not eventually impact the surface somewhere.
What this means, when considering the system as a whole, is that as the solar wind approaches the magnetic 
anomaly, the electrons will begin to be deflected or reflected, due to their lower mass (Figure ~\ref{ingenii}b). 
The ions, with their heavier
mass, will continue towards the surface. An electric field will then be created by this charge separation. This electric
field will be in the opposite direction of the incident flow, and will slow the ions as a positive charge will want to move
in the direction of the electric field, thus enhancing the deflection caused by the anomalous magnetic field. 
The electrons will also feel the effects of the electric field and be pulled closer. The net effect though will be stronger
deflection of the ions than when looking at the individual particle tracking alone, 
and those ions that do impact the surface will have a lower velocity than when the influence of the
the electrons is ignored.  This has been observed by Kaguya, measuring both protons and alpha particles being slowed, heated
and reflected by the magnetic anomalies, while the electrons were accelerated towards the surface [\cite{Saito2012}].

We can still use the particle tracking to get an estimate of both the influence of the magnetic field alone (helping to 
gauge how much of a  role the electric fields play in deflecting particles at magnetic anomalies) 
and the level of complexity in the
vector magnetic field data required to explain the fine detail seen in the swirls, from actual swirls patterns 
to dark lanes near the swirls.
For most of the Moon, our only measurement of surface magnetic fields comes from electron reflectometery measurements, which 
only return total magnitude, not vector fields [\cite{Mitchell2008}].

Besides being able to deflect particles for slow solar wind speeds, the results suggest that the Mare Ingenii 
can also deflect nominal and fast solar wind particles. Figures ~\ref{ingenii}f and ~\ref{ingenii}h show the
surface density maps for nominal solar wind speed and a fast solar wind, respectively. The surface density map for
mid-range SEPs is shown in Figure ~\ref{ingenii_seps}a.  Deflection of the faster solar wind 
particles around the strongest portion of the anomalous region still occurs 
(\(\sim 35^{o}\)S and \(160^{o}\)E, of Figure ~\ref{ingenii}h),
but not as effectively as the baseline case, with a density on the order of 10 particles \(\rm{cm}^{-3}\) 
in the main void region, as opposed to approximately 2-4 particles \(\rm{cm}^{-3}\) 
in the baseline case (Figure ~\ref{ingenii}d).
The density of solar wind particles impacting the surface around the strongest region is higher for both 
faster solar wind cases (Figures ~\ref{ingenii}f and ~\ref{ingenii}h) than the baseline case 
as those particles get much closer to the surface before they begin to be deflected, and are thus localized in their impact. 
The density of particles impacting the surface surrounding the secondary anomaly, at \(\sim 25^{o}\)S and \(170^{o}\)E, exhibits
this same characteristic.  As the initial speed of the particles increases from 200 km \(\rm{s}^{-1}\) to 
400 km \(\rm{s}^{-1}\) to 2000 km \(\rm{s}^{-1}\),
the regions with the highest impact densities transitions to regions more adjacent to the anomalies. This is also associated with
the fact that, with increased speed (and those kinetic energy), the particles are being deflected less before they impact 
the surface. This behavior is in agreement with the idea that the dark lanes are locations of increased weathering
as particles are preferentially deflected into those regions [\cite{Kramer2011a}].  

The impact density (Figure ~\ref{ingenii_seps}a) and flux for SEP particles show no apparent influence by the anomalies
on the incident particles, when compared to a run with the same incident velocities but no anomalous 
magnetic field present (not shown). 
Examination of the components of the particle velocity show some influence though by the anomalies.
Figures ~\ref{ingenii_seps}b and ~\ref{ingenii}c show the components 
of the velocity tangential to the surface for SEP particles. The two strongest anomaly regions cause
some deflection in the incident particles by converting some of the kinetic energy directed toward the surface
into kinetic energy parallel to the surface, as is evident by magnitudes parallel to the surface that are non-zero. 
As the velocity components tangential to the surface are, at most, 0.75\% 
of the incident velocity, it is not enough to be noticeable in the density or flux maps.

\subsection{Reiner Gamma} 

\begin{figure*}[t]
\begin{center}
\includegraphics[height=0.8\textheight]{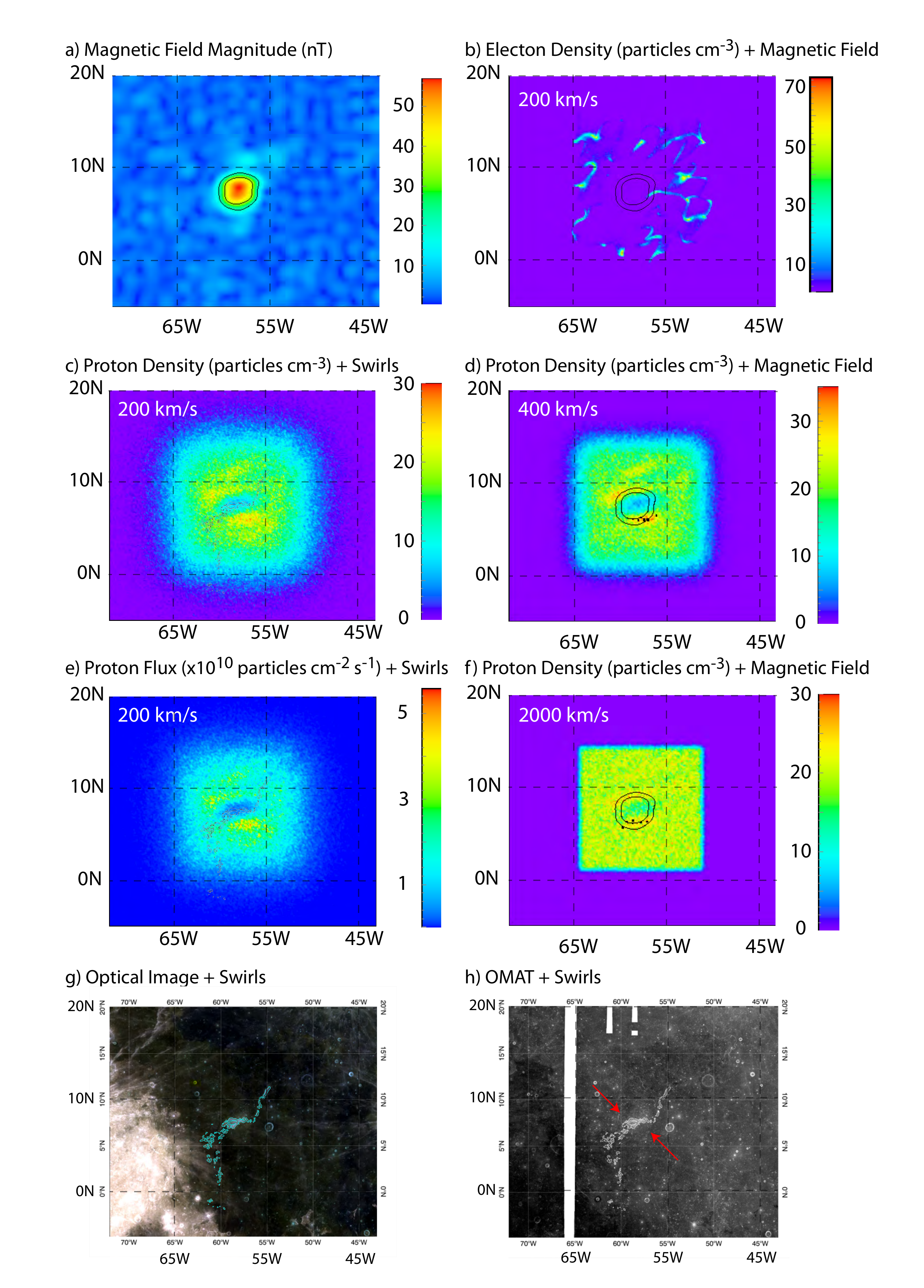}
\caption[Reiner Gamma Maps]
{\textbf{Reiner Gamma} - The (a) surface magnetic field at Reiner Gamma.  Also shown are (b) surface electron density
with magnetic field contours overlayed, 
(c) surface proton density with swirls (in black), and (e) surface proton flux with swirls (in black), 
for the case of particles with a 
mean incident velocity of 200 km \(\rm{s}^{-1}\) and an IMF of \(\rm{B}_{\rm{vertical}} = -2 \rm{nT}\). 
Also shown are surface proton densities with magnetic field overlayed for protons with an incident velocity of
(d) 400 km \(\rm{s}^{-1}\) and (f) 2000 km \(\rm{s}^{-1}\). The swirls are mapped from the (g) optical image
and (h) a spectral image (in cyan). The magnetic contour lines in all four figures show anomalous magnetic field 
magnitudes of 20 nT and 35 nT.
Animations of figure components can be found at: http://earthweb.ess.washington.edu/eharnett/papers/LunarImpactMovies/}
\label{reiner}
\end{center}
\end{figure*}

Figure ~\ref{reiner} shows the results for the Reiner Gamma anomaly region. 
As for Ingenii, the Reiner Gamma swirls are outlined in cyan based on optical imagery in 
Figure ~\ref{reiner}g. These outlines are shown, in black, overlain on the proton density map (Figure ~\ref{reiner}c), 
and proton flux map (Figure ~\ref{reiner}e) for the baseline velocity, to compare model results 
with the locations of the high concentration of swirls.
For this case, spectral imagery indicating optical maturity (OMAT) 
from Clementine is also available and shown in Figure ~\ref{reiner}h, with the swirls marked in white.
For the Reiner Gamma case, there is a reduced density and flux near the strongest portion of the magnetic field 
and an enhanced density and flux surrounding the anomaly (Figures ~\ref{reiner}c and ~\ref{reiner}e). 
The flux to the surface in the region of peak field strength at Reiner Gamma is not close to zero though, like at the
regions of strongest magnetic fields of the Mare Ingenii anomaly. 
Instead the flux at the peak field region at Reiner Gamma
is on the order of \(5 \times 10^{9}\) particles \(\rm{cm}^{-2}\) \(\rm{s}^{-1}\), approximately 1/5th the flux 
when no anomalous field is present. 
The density within this region is reduced by approximately an order of magnitude. Like that for Mare Ingenii,
the regions surrounding the central magnetic anomaly experience an enhanced flux, evident by the regions of yellow and 
orange to the north and south of the strong central magnetic field. This deflection of particles around 
the anomaly also leads to a higher density adjacent to this same region. These densities on the order of 
25 particles \(\rm{cm}^{-3}\) are an enhancement above what is seen when no anomaly is present. 

This behavior of decreased density near the central eye of the Reiner Gamma anomaly 
and an increased flux surrounding the eye, 
is qualitatively similar to that seen for the full particle simulations presented in the companion paper 
[\cite{Bamford2015}], in which the central eye of the Reiner Gamma anomaly was modeled as both a single 
dipole with the moment in three different orientations relative to the surface. 
For the cases with the moment parallel to the surface, 
the enhancement in the density surrounding the eye in the full particle simulations is a factor
of 3-5 times background, as opposed to slightly less than a factor of 2 for the results in Figure ~\ref{reiner}c. 
Another difference is that by using a model of the full anomaly for the particle tracking, density enhancements
are only seen to the north and south of the eye, whereas the assuming a dipole magnetic field for the full particle
simulations results in density enhancements surrounding the entire eye, albeit not symmetrically. 

With a magnitude of 58 nT at the center of the magnetic anomaly, the Reiner Gamma anomaly is weaker than the strongest
anomaly region at Mare Ingenii (at 74 nT) but comparable to the secondary region at Mare Ingenii (at 54 nT). The impact 
density at the center of the secondary region at Mare Ingenii (2 - 5 particles \(\rm{cm}^{-3}\)) is comparable 
to that at the center of the Reiner Gamma anomaly, as is the flux. In this case, the speed of the deflected particles is 
modified only negligibly, while the components of the velocity towards the surface and along the East-West direction show
some modification. Particles are deflected up and to the left or down and to the right. 

The very low albedo of the whole region surrounding the swirl means it is not possible to determine if 
enhanced weathering occurs around the anomaly from the optical images alone, as would be predicted by the particle tracking. 
Figure ~\ref{reiner}h shows the OMAT image. The regions of highest flux and density for the particle 
tracking are noted by the red arrows in Figure ~\ref{reiner}h. Enhanced weathering is not apparent in those regions,
on the same scale as the maturity of the highlands material in the lower left corner. The dark lanes contained within 
the eye of the central anomaly (obscured by the swirl mappings in Figure ~\ref{reiner}h, but visible in
the supplemental images) have an OMAT appearance 
similar to the regions surrounding the central eye of the anomaly.

The behavior of the particles as the speed increases in very similar to that seen for Mare Ingenii. The density of impacting 
particles surrounding the anomaly, for the 400 km \(\rm{s}^{-1}\) case is higher than for the 200 km \(\rm{s}^{-1}\), at
35-40 particles \(\rm{cm}^{-3}\)) (Figure ~\ref{reiner}d). 
With the reduced efficiency of the Reiner Gamma region in deflecting the slowest solar wind speeds 
relative to Mare Ingenii (evident in that even at the location of the strongest fields at Reiner Gamma, 
the flux of particles to the surface is non-zero), it
is not surprising that particle tracking indicates that the Reiner Gamma anomaly has only a modest influence on the
fast solar wind particles (Figure ~\ref{reiner}f). 
The influence of the Reiner Gamma anomaly on the SEP range particles is only apparent when
looking at the components of the velocity (not shown). 
Electrons impacting the surface are confined to surrounding the anomaly region (Figure ~\ref{reiner}b).
The efficiency of the Reiner Gamma anomaly in deflecting electrons is indicated by the low peak densities, 
suggesting this region would significant would experience even greater deflection of the protons from the surface.

\subsection{Gerasimovich}

\begin{figure*}[t]
\begin{center}
\includegraphics[height=0.8\textheight]{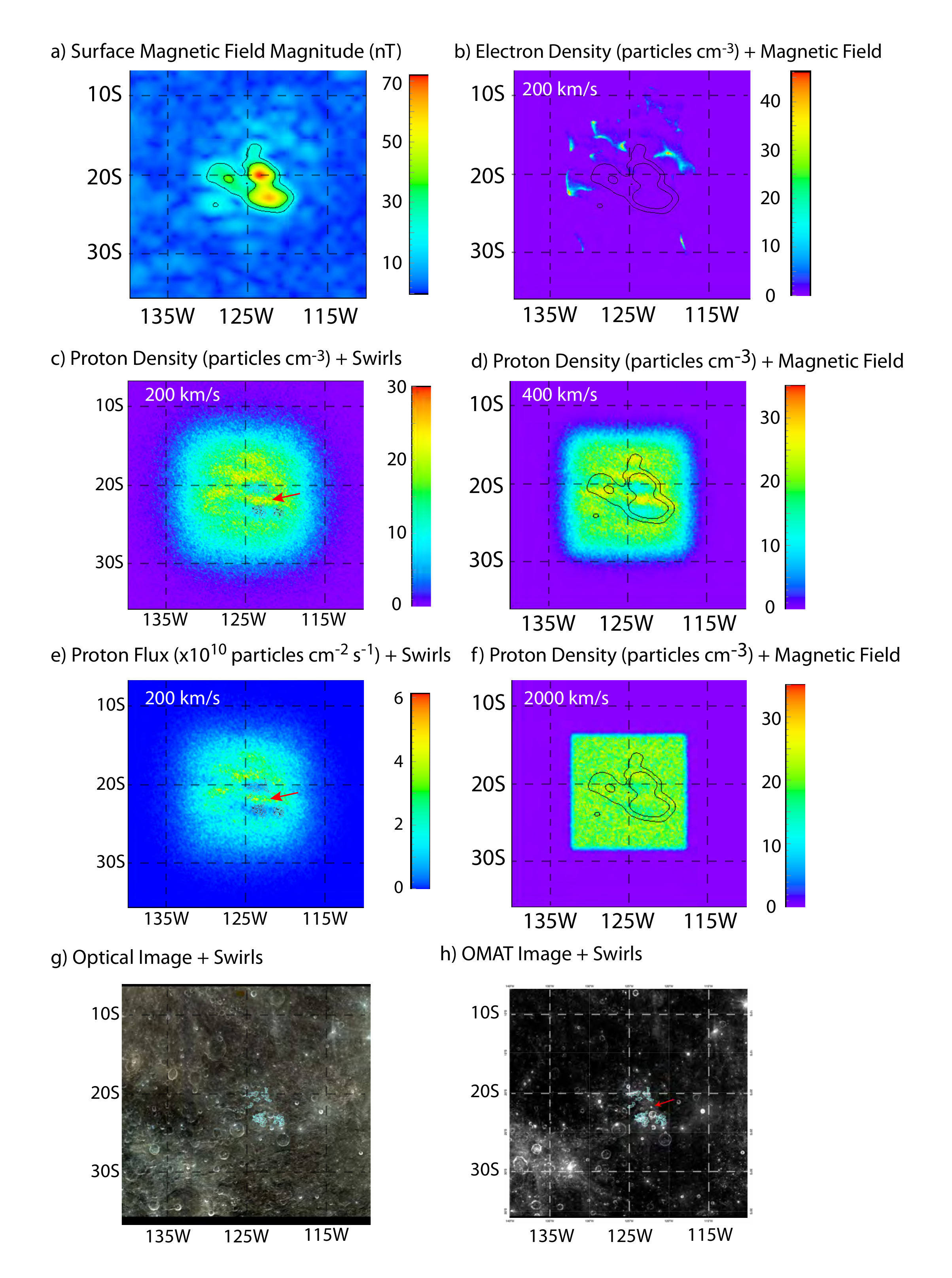}
\caption[Gerasimovich Maps]
{\textbf{Gerasimovich} - The (a) surface magnetic field at Gerasimovich. Also shown are (b) surface electron density
with magnetic field contours overlayed, (c) surface proton density with swirls (in black) overlayed and the (e) surface
proton flux with swirls (in black) overlayed, for the case of particles with a 
mean incident velocity of 200 km \(\rm{s}^{-1}\) and an IMF of \(\rm{B}_{\rm{vertical}} = -2 \rm{nT}\). 
Also shown are surface proton densities with magnetic field overlayed for protons with an incident velocity of
(d) 400 km \(\rm{s}^{-1}\) and (f) 2000 km \(\rm{s}^{-1}\). The swirls are mapped from the (g) optical image
and (h) an OMAT image (shown in cyan). The magnetic contour lines in all four figures show anomalous magnetic field 
magnitudes of 20 nT and 35 nT. The red arrow on (h) indicates the location of a fresh impact crater between the 
two magnetic anomalies that is obscuring what would otherwise be a darker region (meaning higher maturity) 
in the OMAT image. The same region is indicated by the red arrows on (c) and (e).
Animations of figure components can be found at: http://earthweb.ess.washington.edu/eharnett/papers/LunarImpactMovies/}
\label{geras}
\end{center}
\end{figure*}

The particle tracking results show that the Gerasimovich magnetic anomaly is also able to deflect 
incident protons, but not as effectively
as Mare Ingenii or Reiner Gamma (Figure ~\ref{geras}). The format of Figure ~\ref{geras} is similar to 
Figure ~\ref{reiner} with both optical (Figure ~\ref{geras}g) and spectral information from Clementine available 
(Figure ~\ref{geras}h). The swirl contours are shown in cyan on Figures ~\ref{geras}g and ~\ref{geras}h, and in 
black on Figures ~\ref{geras}c and \ref{geras}e. At Gerasimovich, the flux at the 
region of strongest magnetic field is on the order of \(5-7 \times 10^{9}\) particles \(\rm{cm}^{-2}\) \(\rm{s}^{-1}\). 
Similarly, the density at the locations of peak magnetic field is 
reduced, but to a lesser extent than at Mare Ingenii, as it does not approach zero. 
The density of approximately 7-10 particles \(\rm{cm}^{-3}\) is
only 50-60\% that of when no anomalous magnetic field is present. 
Using observations of energetic neutrals coming from the lunar surface made by the Chandrayaan-1 mission, 
\cite{Volburger2012} estimated the shielding 
efficiency the Gerasimovich anomaly to be between 5\% and 50\% over regions with magnetic field strength 
at 30 km between 5 nT and 13 nT, respectively, at low solar wind dynamic pressures. The reduction in surface flux
over the strongest portions of the Gerasimovich anomaly seen in the particle tracking is thus comparable to 
observed the upper end of observed values.  

That the minimum density seen at the surface is greater than that seen for the Mare Ingenii cases 
can not be explained by surface magnetic field strength alone. The peak magnetic 
field strengths at Gerasimovich are comparable to those at Mare Ingenii, where the strongest anomalous magnetic field
resulted in a near zero flux of particles at the location of peak magnetic field. The peak magnetic field strength  
at Gerasimovich is also approximately 20\% stronger than the peak field strength at Reiner Gamma,
which has a comparable flux and slightly lower density at the center of the anomaly.

The swirls at Gerasimovich are strongly co-located with the peak magnetic 
field strengths and minimums in the impact density and flux
(Figures ~\ref{geras}c and ~\ref{geras}e). 
In the OMAT image for Gerasimovich (Figure ~\ref{geras}h), an extended, diffuse region can be observed 
surrounding the swirls (one portion of which is indicated by a red arrow). 
The lighter coloring in OMAT indicates that the diffuse region is of higher maturity 
than the swirls, but lower maturity than the broader, surrounding area.
This region matches the shape of the extended magnetic field
(Figure ~\ref{geras}a) and is surrounded by regions of high impact density and flux (with a red arrow indicating those
region in-between the two main swirl groups in Figures  ~\ref{geras}c and ~\ref{geras}e).  
The high flux and density region between the two peaks in the magnetic field 
magnitude is coincident with a dark region in the 
OMAT map, which would likely appear even darker were it not for the occurrence of a fresh impact crater 
in that location (also red arrow in Figure ~\ref{geras}h).

The reduced efficiency deflecting the slowest particles translates to a reduced influence on faster incident particles.
Gerasimovich can deflect 400 km \(\rm{s}^{-1}\) (Figure ~\ref{geras}d) but the peak densities adjacent to the anomaly, 
at 30 particles \(\rm{cm}^{-3}\), are the smaller than for Mare Ingenii and Reiner Gamma. 
With correspondingly higher densities
in the anomalous regions, this means that the particles are experiencing less deflection by the magnetic field.  
Gerasimovich has little influence on the impact density or flux for the fast solar wind case (Figure ~\ref{geras}f), and 
no influence on the impact density or flux for the SEP case. Like the other anomalies though, modification in the 
velocity components parallel to the surface does occur for both the fast solar wind and SEP cases, but not to the same
extent as more efficient anomalies. 

\subsection{Marginis}

\begin{figure*}[t]
\begin{center}
\includegraphics[height=0.8\textheight]{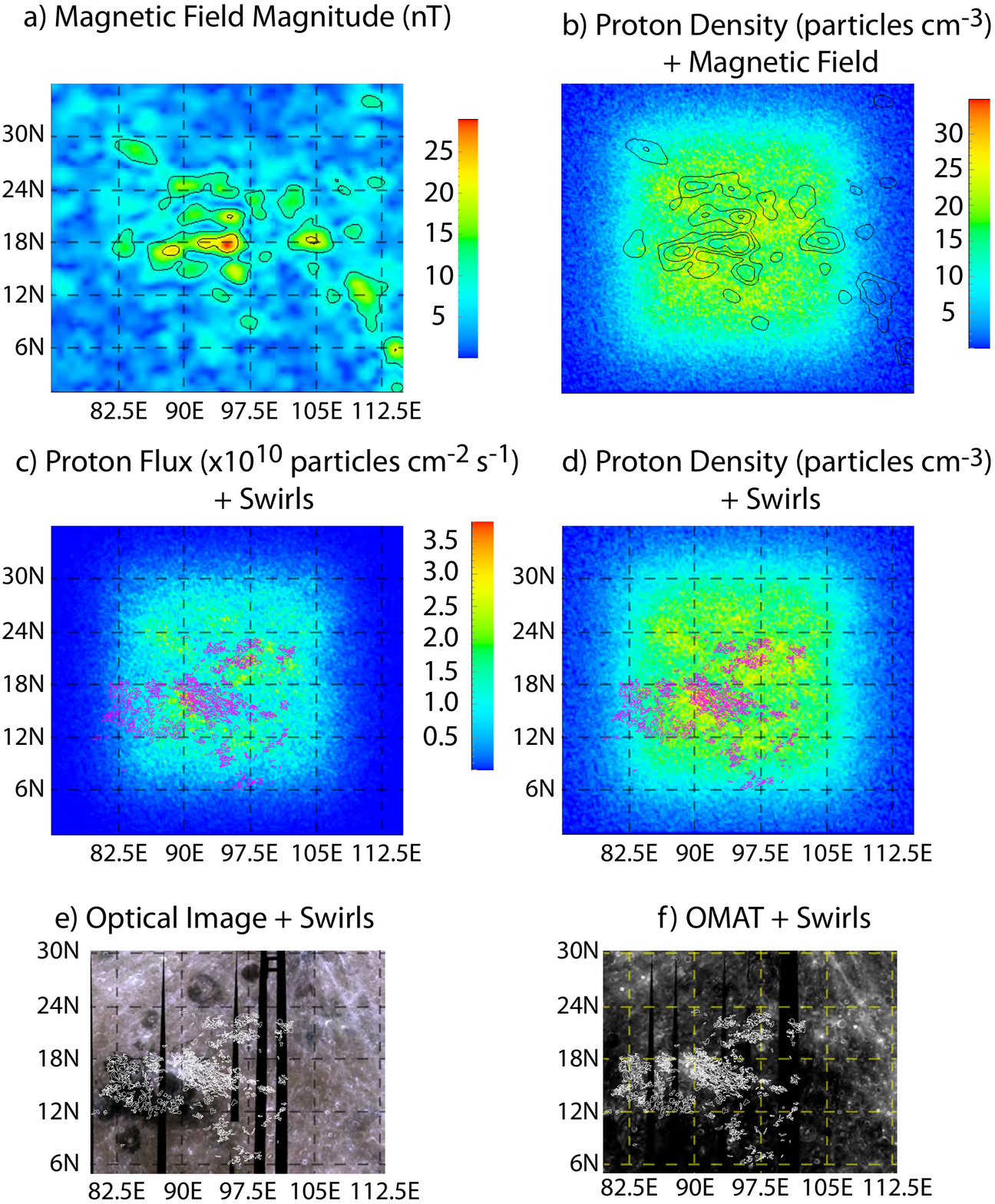}
\caption[Marginis Maps]
{\textbf{Marginis} - The (a) surface magnetic field at Marginis. Also shown are (b) surface proton density
with magnetic field contours overlayed, (c) surface proton density with swirls (in magenta) overlayed and the (e) surface
proton flux with swirls (in magenta) overlayed, for the case of particles with a 
mean incident velocity of 200 km \(\rm{s}^{-1}\) and an IMF of \(\rm{B}_{\rm{vertical}} = -2 \rm{nT}\). 
The swirls are mapped from the (g) optical image
and (h) an OMAT image (shown in white). The magnetic contour lines in figures (a) and (b) show anomalous magnetic field 
magnitudes of 10, 15, 20 and 35 nT. }
\label{marg}
\end{center}
\end{figure*}

With a peak surface magnetic field magnitude of approximately 30 nT, the Marginis anomaly is the weakest of the five
regions investigated in this study. The results  is that while, particles are deflected, the anomaly is much less 
effective in doing so. For all of the other four regions studied, at least some portion of the anomalous region experiences
a surface flux and density of at least half that compared to an unaltered region. This is not true for Marginis.
Figure ~\ref{marg} shows the results for the slow (200 km \(\rm{s}^{-1}\)) particle case as compared to the optical and 
OMAT measurements. For Marginis, the difference 
in peak densities and reduced densities is a factor of two. For the flux, the relative difference is also a factor of
two. The reductions are only a by a factor of 1.3, as compared to when no anomalous magnetic field is present.
The reduced surface density and flux are co-located with regions of magnetic field strengths in excess of 15 nT. 
Only the X component of the velocity of the incident particles shows any modification by the anomalies. This 
component is perpendicular to the largest tangential component of magnetic field. 
Because of the inefficiency of the region in deflecting the slow solar wind plasma, results for higher incident plasma
speeds are not shown. 

Mare Marginis is a region of significant interest because of its prominent lunar swirls both on
mare and highland soils. Marginis was chosen for this study to compare results of a similar region of 
complex swirl patterns, Mare Ingenii, which was also analyzed in \cite{Kramer2011a}. It cannot be ignored 
that the pattern of swirls at Marginis appear to emanate from Goddard A, a fresh 11 km crater, 
which ejected bright highlands material over the surrounding mare and highlands regions.
The initial mapping of swirls at Marginis also used the quasi-slope map generated from the 
LRO WAC 643 nm normalized reflectance map and the LRO GLD100 map. 
The intricate pattern of the swirls at Marginis necessitated the use of
LRO’s Narrow Angle Camera (NAC) [\cite{Robinson2010}] in some locations to map the swirls in detail. Although some
swirls continue in the highlands east of this region, mapping was restricted to the mare and nearby highlands
for the purpose of this study (Figures ~\ref{marg}e and ~\ref{marg}f).

Outlines of the mare and swirl overlaid on both the surface flux (Figure ~\ref{marg}c) and 
density (Figure ~\ref{marg}d). 
Swirls were found across the highlands and some of the mare regions, in locations where the particle
tracking shows a decreased proton flux. Regions of increased proton flux tended to be lacking in
swirls, while regions of decreased proton flux tended to be rich in swirls. The swirls were quite obvious in
the regions of reduced proton flux in both the high-FeO mare and low-FeO highlands. 
Locations of significant interest are
Goddard basin (at 14.8N and 89.0E) and Ibn Yunus basin (at 14.1N and 91.1E), 
both of which are vast regions that lack any observable swirls despite being within the 
same radial distance from Goddard A as other locations that exhibit prominent swirl patterns
(Figure ~\ref{marg}e). The particle tracking predicts a high
proton flux across both mare-filled basins, which explains the lack of swirls within the basins (Figure
~\ref{marg}c). The target mare soils are rich in FeO with which to create nanophase iron, 
causing the Goddard and Ibn Yunus basin
floors to be weathered more efficiently than the nearby highlands. This effect is observable in the 
OMAT map (Figure ~\ref{marg}f)
where Goddard and Ibn Yunus appear dark (i.e. mature), while the surrounding highlands appear brighter
(i.e. immature) and swirled. Although some of its ejecta is confused with the swirls, across Goddard 
and Ibn Yunus basins, as well as a few locations in the highlands, where Goddard A ejecta can be observed to 
have a typical impact ejecta pattern, that is, not swirled, and the ejecta appear more mature than radially 
proximate swirls.
This is strong supporting evidence that the ejecta, which landed where the simulations predict the proton flux is high, 
are being weathered at an accelerated rate while those that landed on highland swirls are being preserved,
further supporting the solar wind magnetic deflection model, even at this weaker anomaly.

When comparing the optical images for Ingenii and Marginis, it would be
tempting to conclude that the magnetic fields would be similar, based upon a comparison of swirl extent and contrast
with surrounding terrain. As the above analysis shows though, it is difficult to explain the swirls based upon the 
limited efficiency of the magnetic field alone in deflecting incident particles, as compared to other anomalies.  

\subsection{Northwest of Apollo}

\begin{figure*}[t]
\begin{center}
\includegraphics[height=0.8\textheight]{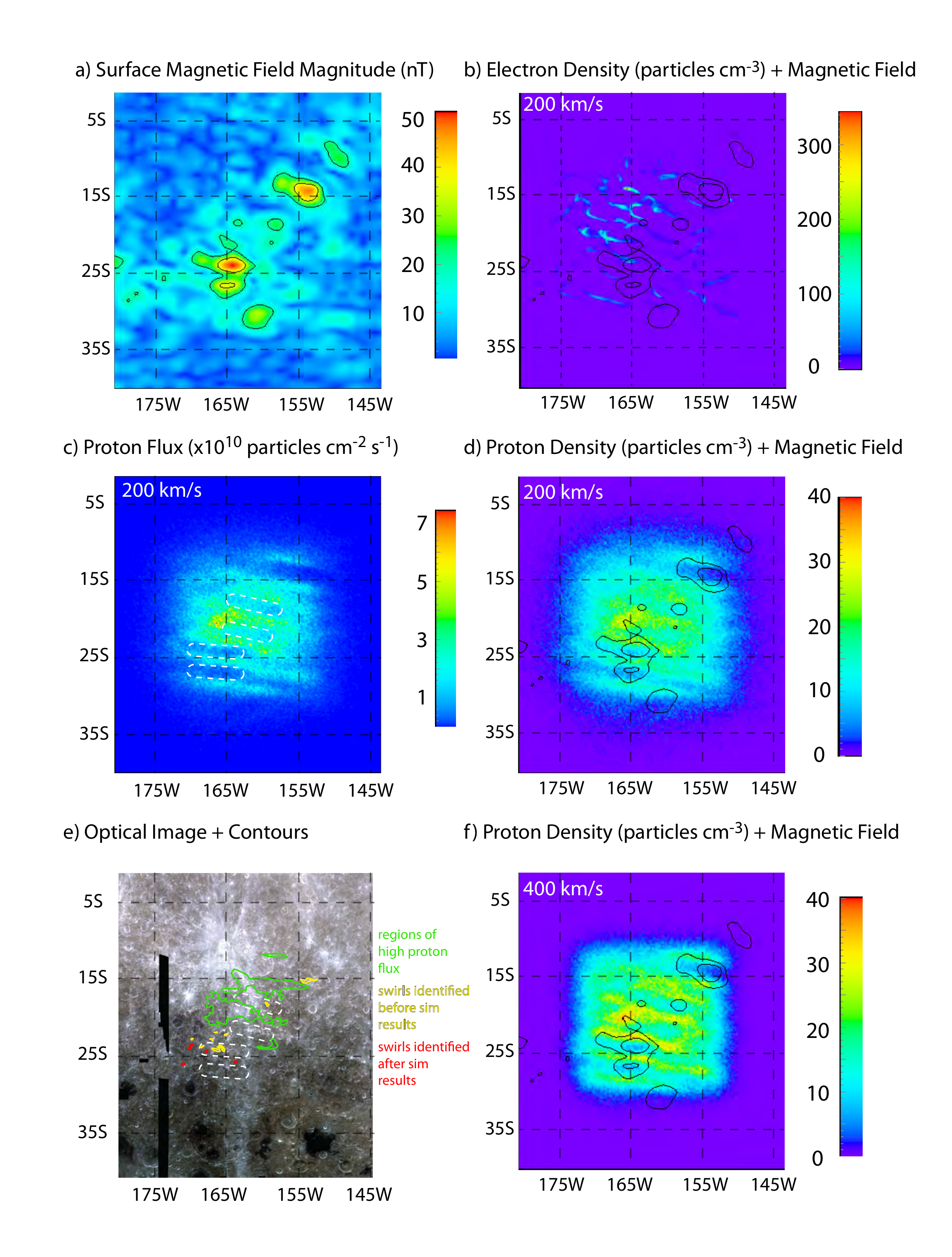}
\caption[NW of Apollo Maps]
{\textbf{NW of Apollo} - The (a) surface magnetic field at Northwest of Apollo. Also shown are (b) surface electron density
with magnetic field contours overlayed, (c) surface proton flux, and (d) surface proton density with magnetic field
contours for the case of particles with a 
mean incident velocity of 200 km \(\rm{s}^{-1}\) and an IMF of \(\rm{B}_{\rm{vertical}} = -2 \rm{nT}\). 
Regions of low proton flux are mapped from (c) for comparison with the (e) optical image and indicated by white ovals.
The surface proton density for an incident velocity of 400 km \(\rm{s}^{-1}\) is shown in (f).
The contour lines in both figures show anomalous magnetic field magnitudes of 20 nT and 35 nT.
The (e) optical image also shows regions of high proton flux (green contours), high proton density (red contours) and 
strong magnetic field (blue contours).  Regions of low proton flux from (c) are shown also shown in (e).  }
\label{nw}
\end{center}
\end{figure*}

\begin{figure}[t]
\begin{center}
\includegraphics[height=0.8\textheight]{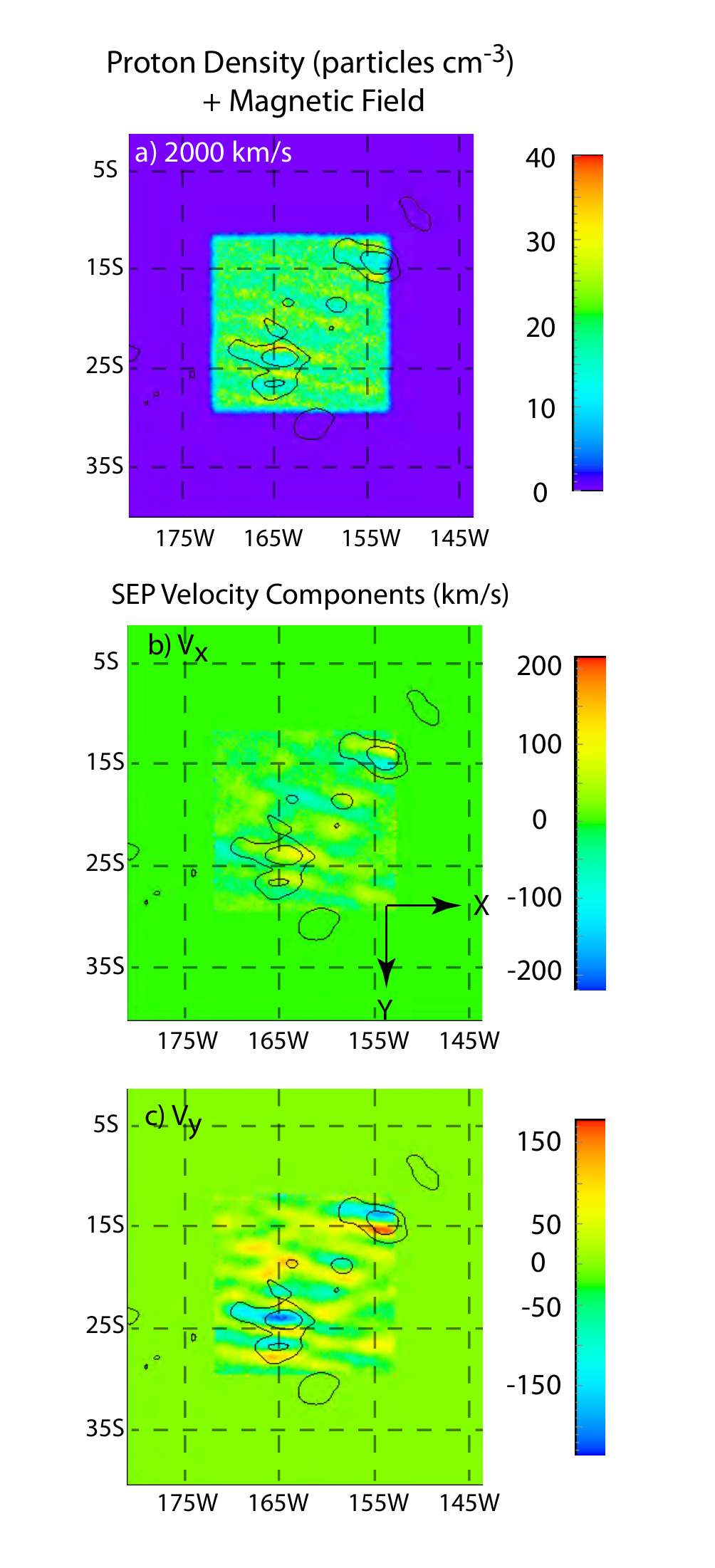}
\caption[NW Apollo Velocity]
{\textbf{NW of Apollo} - The (a) surface density at NW of Apollo for an incident velocity of 2000 km \(\rm{s}^{-1}\). 
Also shown are average components of the velocity of the SEP (40,000 km \(\rm{s}^{-1}\))  
particles when they impact the surface (b) along the
East-West direction and (c) along the North-South direction at NW of Apollo. Arrows indicate the positive
directions for each component. The magnetic field contours on all three images are identical to those in ~\ref{nw}.}
\label{nw_seps}
\end{center}
\end{figure}

The NW of Apollo region is comprised of two anomalies with comparable surface magnetic field strength. 
Both anomalous regions lead to reduced density and flux when compared to the case with 
no magnetic field present. The anomaly in the upper right corner (\(\sim 15^{o}\)S and \(155^{o}\)W) (Figure ~\ref{nw}a), 
while weaker in total magnetic field magnitude than the anomaly at \(\sim 25^{o}\)S and \(165^{o}\)W, 
is more efficient in reducing the flux of particles to the surface (Figure ~\ref{nw}c). 
In this region the particle density is approximately
75\% less than when no magnetic field is present, as opposed to the 60\% reduction for the region at the center of the
strongest anomaly. The components of the magnetic field show that the North-South components of the magnetic field at the 
secondary anomaly are 1.5 to 2.0 times larger than the North-South components for the strongest anomaly. 
The East-West components at the primary and secondary anomaly are comparable. The stronger magnetic field 
magnitude at the primary anomaly comes from  a larger vertical component. This may partially explain why that 
portion of NW of Apollo region is less effective in deflecting the 
incoming solar wind. The east-west component of the magnetic field for the strongest anomaly in the Mare Ingenii region
are comparable in magnitude to the radial component. This aspect will be discussed further in the next section. 

The high albedo of the ejecta rays from the crater Crookes, to the north of the anomaly region, make
swirl identification difficult from the optical images, but
the particle tracking can assist with refining the regions to look at. 
The white ovals in Figure ~\ref{nw}c mark the regions of low impact flux. They are replicated in the optical image 
(Figure ~\ref{nw}e).  The albedo inside the two ovals between 
\(\sim 20^{o}\)S and \(25^{o}\)S does appear to be lighter than the surrounding regions. That this higher albedo
is perpendicular to the ejecta rays suggests it may be associated with the anomalous region instead. 

The extended structure of the magnetic field throughout the region means the density of particles impacting the surface
is more complex than the previous cases. The near equivalent magnitude of the primary and secondary anomalies deflects
particles in toward the region between the two, but a weak anomaly near the center of the region is still strong 
enough to lead to moderate deflection. Thus the highest densities and fluxes are in the region straddling the three
anomalies. This structure holds for the faster solar wind cases as well (Figure ~\ref{nw}f and Figure ~\ref{nw_seps}a) 
It is also visible in the modification of the tangential velocity components for the SEP case (Figures ~\ref{nw_seps}b
and ~\ref{nw_seps}c) but not the density or flux. 

The electrons show more deflection than protons with a similar speed, but are still able to impact near the
central portion of the anomalous region, in between the two regions of strongest magnetic fields
 (Figure ~\ref{nw}b). The surface density of impacting electrons is also the highest for NW of Apollo, when 
compared to the previous three anomalies, investigated. 
That the electrons can access the central portion of NW of Apollo
is likely associated with the more complicated magnetic field, which is also manifest in the complex 
impact patterns seen for the protons. With more access by the electrons, the
electric fields generated by the charge separation between protons and electrons (which is a function of the separation
distance) should be smaller than for regions where the particle tracking shows little access by the electrons. The protons
will be slowed less by the smaller electric field, potentially increasing access to the surface.

Swirls in NW of Apollo have been identified (e.g. \cite{Blewett2011}), but had not been mapped,
likely because this heavily cratered highlands region has little contrast between swirl and background albedo,
and the complicated topography causes albedo anomalies across the region. Mapping swirls for NW Apollo
proved more difficult than other regions so we generated a quasi-slope map by contrasting the LRO WAC 643 nm 
normalized reflectance map with the LRO GLD100 topographical map [\cite{Scholten2012}]. 
This provided high albedo swirls. 
Even so, only about half of the swirls shown here were found with this method. The rest were found using
particle tracking as a guide, doubling the number of identified swirls. All of the swirls mapped using the 
combined set of techniques are shown in in Figure ~\ref{nw}e. This highlights that quick particle tracking 
(as opposed to more computationally expensive full particle simulations) can be a useful tool in helping refine
a search area when mapping swirls at other anomalies in which swirl identification is complicated by the 
surrounding material.

\section{Discussion and Conclusions}

\begin{figure*}[t]
\begin{center}
\includegraphics[width=0.95\textwidth]{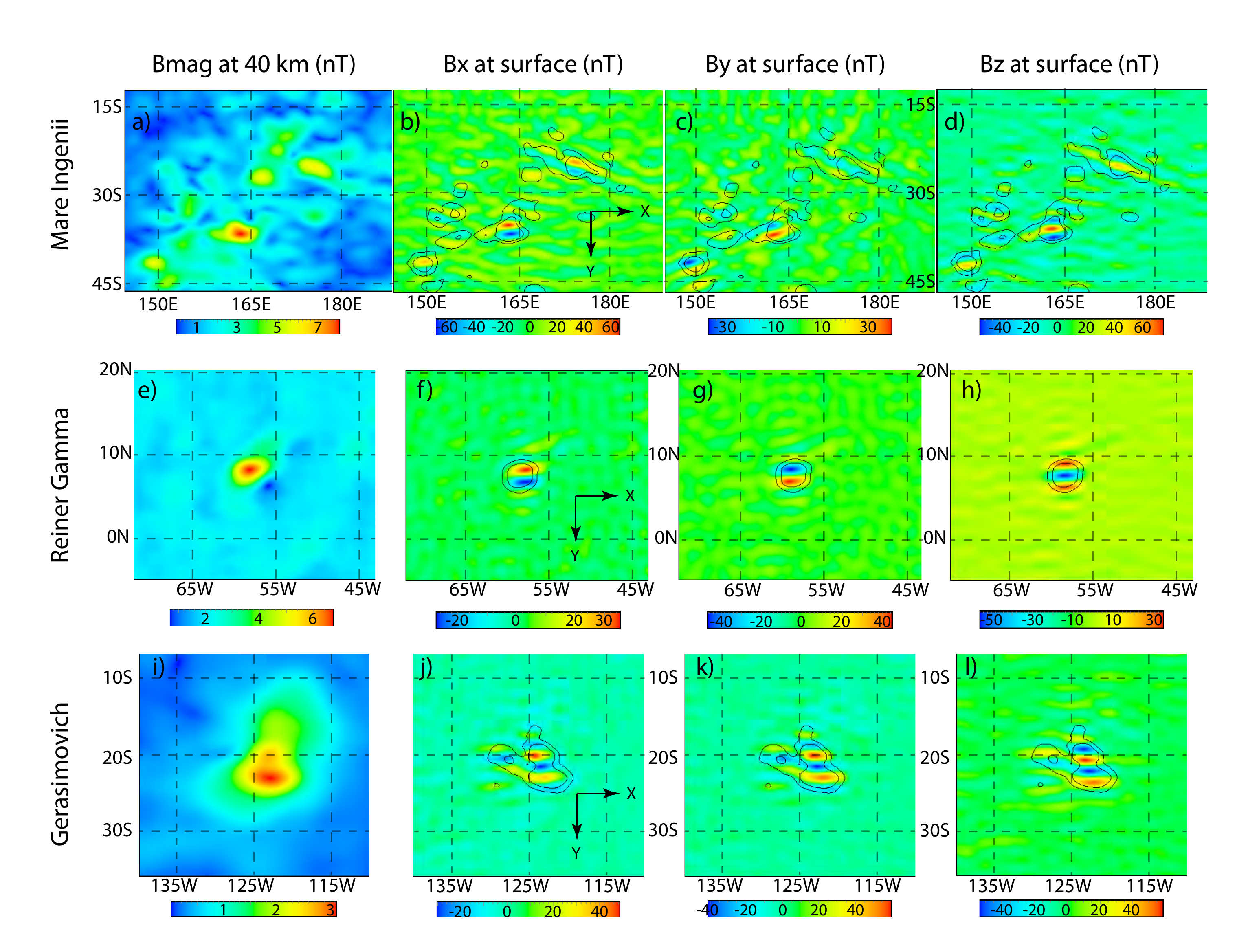}
\caption[Anomaly Coherence]
{The magnitude of the magnetic field at 40 km (first column) for the Mare Ingenii, Reiner Gamma and 
Gerasimovich anomalies and the magnitude of the 
Bx (second column), By (third column), and Bz (last column) components of the magnetic field at the 
surface. Magnetic field contour lines are only shown for 20 and 35 nT.}
\label{surfmag}
\end{center}
\end{figure*}

\begin{figure*}[t]
\begin{center}
\includegraphics[width=0.95\textwidth]{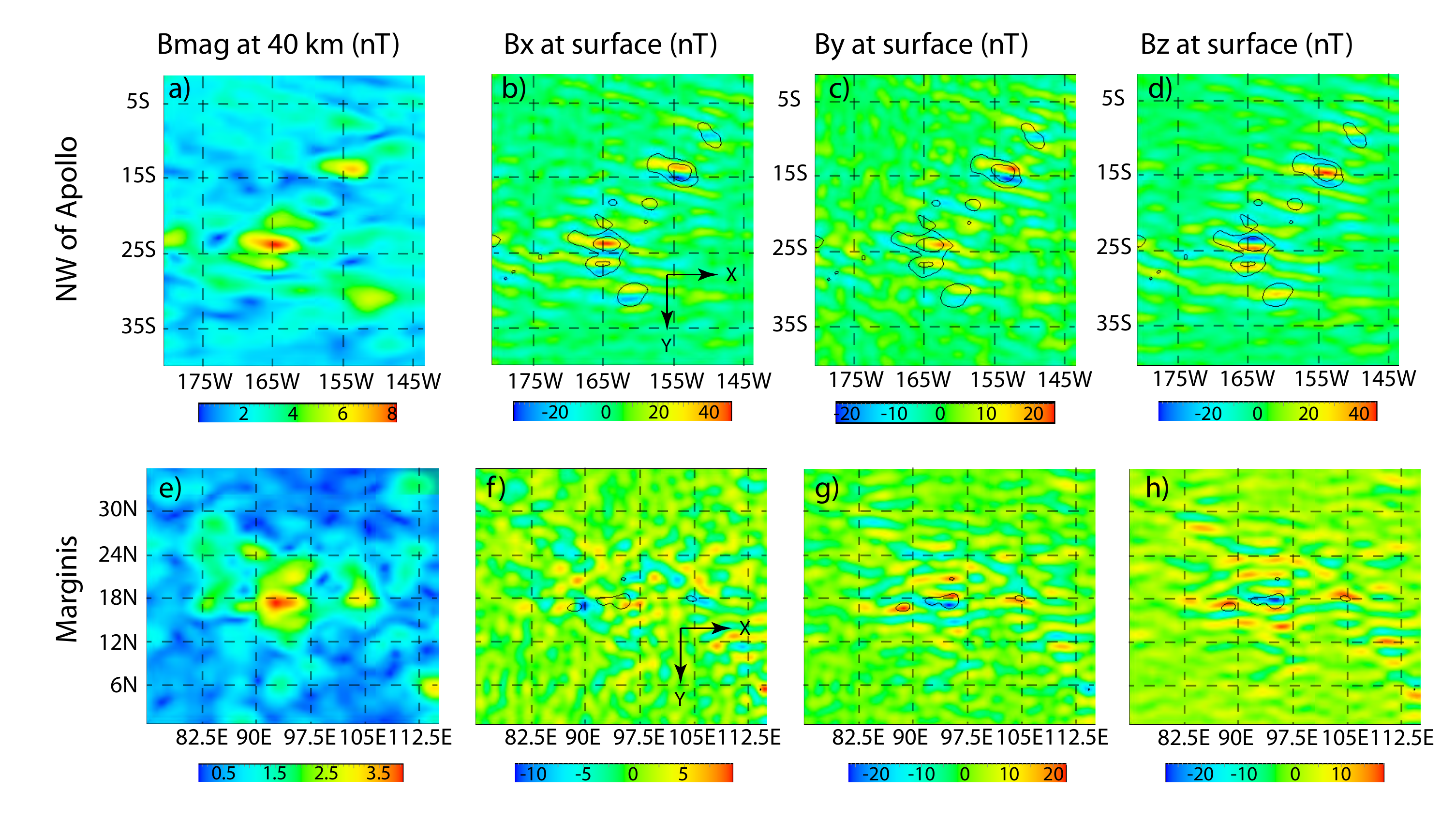}
\caption[Anomaly Coherence pt 2]
{The same format as Figure ~\ref{surfmag} for NW of Apollo and Marginis.  
For consistency, magnetic field contour lines are only shown for 20 and 35 nT, even at the weak anomaly Marginis.}
\label{surfmag2}
\end{center}
\end{figure*}

One problem in understanding the plasma physics occurring near the swirls from observations is that it is not possible
to deconvolve all of the different processes to understand the relative importance of each, and ultimately resolve
the origin of the swirls - is it deflection of the weathering solar wind or the transport of charged dust 
by the subsequent electric fields \cite{GarrickBethell2011}, or are both occurring? This has left open questions 
with regard to how important the electric fields are in deflecting incident plasma and if the anomalous magnetic
field alone can deflect ions, due to the small scale size of the anomalous regions relative to the proton
gyroradius. By using particle tracking only, the results presented in this paper can shed light on how effective
the anomalous magnetic field alone is in deflecting incident protons of varying energies and what aspects of that
anomalous magnetic field are most important.  

The results from this study further highlight that the small-scale size of the lunar magnetic anomalies do not prevent them
from at least partially deflecting incoming solar wind particles, from slow to fast solar wind speeds. And while none
of the anomaly regions investigated could deflect moderate SEP-energy particles in such a way produce local impact density
variations, all of the anomalies could influence the velocities of moderate SEP particles. The effectiveness of each 
anomaly region in deflecting particles does not scale exactly with peak surface magnetic field strength, as one might expect
from magnetic mirroring. It is important to note that while the  process of magnetic mirroring does not depend on the 
direction of the converging magnetic field, just the magnitude, 
it does assume that the scale size of the magnetic mirror region is larger than the gyro-radius of the incident particle
(which is a function of the kinetic energy of the incident particle).

The peak surface magnetic field strengths on the order of 50-70 nT, correspond to a gyro-radius of 30-40 km for the slowest 
speed investigated and 300-400 km for the fast solar wind case. A magnetic field strength of 25 nT corresponds
to a gyro-radius of 84 km for the slowest case and 1670 km for the fast solar wind case. 
The anomaly region most effective in deflecting particles at Mare Ingenii, has magnetic field 
25 nT or greater spread out over approximately 40 km by 30 km. The secondary anomaly at Mare Ingenii, 
while weaker in peak magnitude, has magnetic field 25 nT or greater spread out over approximately 
55 km by 30 km.  While Gerasimovich has a peak surface magnetic field strength comparable to Mare Ingenii, 
the anomalous magnetic field is more localized to single anomaly region approximately 50 km by 40 km. The Reiner Gamma
region is the most localized, with a region of magnetic field 25 nT or greater that is circular in shape, with 
a diameter of approximately 20 km. The local anomaly in the NW of Apollo region that is most effective in deflecting solar 
wind particles has magnetic field 25 nT or greater over a 30 km by 25 km region. The other anomaly in the region, 
that is less effective in deflecting particles, is larger at approximately 50 km by 35 km. Marginis has a region
of magnetic field on the order of 15 nT spread out over approximately 60 km by 50 km, but only a very small area 
exceeding 25 nT. 

That 1) the effectiveness of Gerasimovich in deflecting particles is comparable to Reiner Gamma, 2) the 
larger anomaly in NW of Apollo is less effective than the smaller anomaly, and 3) Marginis can deflect
particles at all with sufficient efficiency to result in a complex region of swirls, all indicate that another
quality of the magnetic field, namely coherence, is important as well. 
The coherence of the magnetic field can be thought of as a measure of how much the magnetic field changes orientation 
over a given spatial distance. For example, Reiner Gamma has a peak surface
magnetic field strength $\sim$ 80\% that of Gerasimovich but is highly localized and the region is circular in shape.  
The size of the gyro-radius of the incident particles will sense both the coherence
of the anomalous magnetic field and the scale size of the magnetic anomaly, as an incoherent anomaly region will
not have large regions of converging magnetic field, due to the field changing orientation over small distances.

Coherence can be assessed by looking at the coefficients of the spherical harmonic expansion used to
model the anomalous magnetic field. The larger the dipole term is relative to the higher order terms indicates a more 
coherent magnetic field. Unfortunately the model used to recreate the anomalous magnetic field did 
not allow for this type of analysis.
Instead other aspects of the magnetic field were used to estimate how coherent the magnetic field is in
each region - fall off with altitude and component analysis (Figures ~\ref{surfmag} and ~\ref{surfmag2}). 

The higher the order the moment in a spherical harmonic expansion, the faster the field, from that term, 
decreases with distance from the source. The first column in Figures ~\ref{surfmag} and ~\ref{surfmag2}  
shows the magnitude
of the anomalous magnetic field in each region at 40 km. The analysis of the impact maps indicates that Mare
Ingenii is the most effective in deflecting incoming solar wind, Marginis and Gerasimovich the least effective, 
with Reiner Gamma and NW of Apollo somewhere in between. The ranking of NW of Apollo depends on which portion 
of the anomalous region
one looks at. This ranking can be partially explained by looking at the magnetic field above the surface. While 
Gerasimovich has some of the largest surface magnetic field strengths, 
the magnetic field for Gerasimovich is the weakest at 40 km.
And while Marginis has the lowest surface magnetic field magnitudes, the field strength at 40 km is still
15\% that of the surface field strength (for comparison, the peak magnitude at 40 km above Gerasimovich is 
5\% that of the peak surface field strength). 
The fast fall-off of the field at Gerasimovich, helps explain why the anomaly is not as effective 
as Reiner Gamma, in deflecting incoming solar wind, even though the surface field strength at Gerasimovich is
stronger. 
NW of Apollo, in contrast, has the strongest fields at 40 km, even though it has the second weakest surface
field of the five regions studied. To explain why it is not the most effective in deflecting particles, and why
Reiner Gamma seems more effective than it should be given its weak magnetic field strength at 40 km, requires
looking at the details of the magnetic field.

When looking at the components of the magnetic field, a perfect dipole with the moment
perpendicular to the surface would look very similar to the images also for Reiner Gamma 
(Figures ~\ref{surfmag}f - ~\ref{surfmag}h) - radial magnetic field exiting at the center, with oppositely 
directed radial field at the edge (Figure ~\ref{surfmag}h), and anti-symmetric tangential magnetic 
fields centered about the radial magnetic field (Figures ~\ref{surfmag}f - ~\ref{surfmag}g). The very 
dipole-like nature of the Reiner Gamma anomaly helps explain why its effectiveness in deflecting particles
is similar to Gerasimovich, even through its peak surface field strength is weaker. It also explains why the
field magnitude at 40 km above Reiner Gamma is larger than above Gerasimovich.  
Figures ~\ref{surfmag}b - ~\ref{surfmag}d show that
the strongest anomaly in the Mare Ingenii region has dipole-like characteristics while the field within Gerasimovich
is more complex. At NW of Apollo, only the anomaly around \(\sim 15^{o}\)S and \(155^{o}\)W, is dipole-like, 
and that region is the most effective in deflecting particles.  Thus the more dipole-like (or coherent) the 
anomalous magnetic field (both in terms in the components and the decrease in strength with distance), 
the more effective it is at deflecting particles, when all other aspects are the same.

One of the tangential components (Y) of the magnetic field at Marginis is comparable in strength to 
the same component at Ingenii (Figure ~\ref{surfmag2}g vs. ~\ref{surfmag}c), but the other components at
Marginis are much weaker than those same components at all the other anomalies. 
That the magnetic field at Marginis does not fall off as quickly as Gerasimovich and that the surface magnetic field is
primarily parallel to the surface helps explain the existence of the swirl patterns at Marginis, but not the
extensive nature. 

It is important to note that one issue the results 
highlight is that the resolution of the model magnetic field used in this study
is typically much coarser than the optical images. This becomes most apparent when analyzing 
the Reiner Gamma and Gerasimovich anomalies. The swirl regions in both of these cases  span only a few 
degrees in lateral extent, and corresponds to only ~ 20 grid points in the simulation magnetic field. 
The true resolution of the magnetic field model is comparable to the lowest altitude of the satellite making the observations. 
The data used to generate the magnetic field models used for this work came from Lunar Prospector, 
which produced global magnetic field maps down to 30 km and made local magnetic field measurements down to 20 km
[\emph{M. Purucker}, 2014 private communication]. This means that the actual magnetic field observations are 
even courser than the model magnetic field. The lower resolution of the magnetic field data will act to
average out more complex (i.e. small-scale) structure, making the magnetic field appear to be more coherent 
than it actually is. Recent modeling using low altitude (~10 km) observations by Kaguya \cite{Tsunakawa2015},
in addition to the low altitude Lunar Prospector observations, 
has allowed for the generation of surface vector magnetic field maps with resolution of \(0.2^{o}\) for some regions.
Observable differences on the order of many 10s of nT were seen for the Reiner Gamma region. This still may
not be sufficient to explain some of the smallest swirl features. 

The work presented here shows that impact maps for 3D particle tracking at lunar magnetic anomalies can be
correlated with observations of the the lunar swirls in the same regions. 
Although the small-scale swirl features cannot be matched with the results of the simulations due to the 
disparity in the spatial resolutions of the imaging data and the magnetic field data, the simulations do show 
that protons are consistent with the spatial 
pattern of the swirls; that is, protons are deflected away from the locations of the high-albedo swirls and 
onto inter-swirls or swirl-adjacent locations. This is consistent with the conclusions of 
\cite[]{Kramer2011a,Kramer2011b} that solar wind ions are the dominant agents responsible for the creation of 
nanophase iron, which is largely responsible for the spectral characteristics of space weathering 
and  optical maturation [e.g. \cite{Hapke2001}, \cite{Noble2007}]. On the swirls, 
the decreased proton flux slows the spectral effects of space weathering (relative to non-swirl regions) by 
limiting the nanophase iron production mechanism almost exclusively to micrometeoroid impact vaporization/deposition. 
Immediately adjacent to the swirls, maturation is accelerated by the increased flux of protons deflected 
from the swirls. Our results show that the shape and strength of the magnetic anomalies, independent of 
an induced electric field, can explain the deflection and focusing of incident protons at solar wind velocities 
at the distance of the Earth-Moon system. Although this may not fully represent the intricacies of the interaction, 
it is an important result for understanding and further refinement of relevant models.

More work needs to be done though
to correlate the small-scale details of the swirls and dark lanes with impact maps. 
While the next step in the process involve conducting fully self-consistent 3D particle simulations of the
solar wind interacting with realistic anomalous magnetic fields, the above analysis
suggests that may not be enough to explain the details of the features seen at swirls. A companion
paper [\cite{Bamford2015}] shows results from full 3D particle simulations using realistic proton to 
electron mass ratios for both a single dipole and a double dipole system. 
Due to the heavy computational load associated with such simulations, 
only the central portion of the Reiner Gamma region was investigated, but this technique works well
for the central portion of the Reiner Gamma region as the anomalous magnetic field appears to be very 
coherent and dipole-like. The above analysis indicates that even with full particle simulations of more
extended regions around anomalies, the results will most likely show a disconnect with the intricate features 
seen at swirls until much higher resolution vector magnetic field measurements are made near the surface. \\\\\\

\section{Acknowledgments}
The authors would like to thank Dr. Michael Purucker and Dr. Joseph Nicholas for providing us with the 
source code to calculate the lunar magnetic fields. 
This research was supported by the NASA LASER grant \#NNX12AK02G.


 \bibliographystyle{acmtrans-ims}
 \bibliography{moonref.bib}

\end{document}